\pdfoutput=1
\documentclass[12pt,a4paper]{article}
\usepackage{float}

\usepackage{ifthen} % for conditional statements
\newboolean{pdflatex}
\setboolean{pdflatex}{true} % False for eps figures 

\newboolean{articletitles}
\setboolean{articletitles}{true} % False removes titles in references

\newboolean{uprightparticles}
\setboolean{uprightparticles}{false} %True for upright particle symbols

\newboolean{inbibliography}
\setboolean{inbibliography}{true} %True once you enter the bibliography

\def\paperauthors{LHCb collaboration} % Leave as is for PAPER and CONF
\def\paperasciititle{LHCb paper} % Set ASCII title here
\def\papertitle{Bose-Einstein correlations of same-sign charged pions in the forward region in $pp$ collisions at \mbox{$\sqrt{s}$ = 7 TeV}} % Latex formatted title
\def\paperkeywords{{High Energy Physics}, {LHCb}} % Comma separated list
\def\papercopyright{CERN on behalf of the LHCb collaboration}
\def\paperlicence{CC-BY-4.0}
\def\paperlicenceurl{https://creativecommons.org/licenses/by/4.0/}

\usepackage[top=1in, bottom=1.25in, left=1in, right=1in]{geometry}

\usepackage{microtype}
\usepackage{lineno}  % for line numbering during review
\usepackage{xspace} % To avoid problems with missing or double spaces after
\usepackage{caption} %these three command get the figure and table captions automatically small

\usepackage{graphicx}  % to include figures (can also use other packages)
\usepackage{color}
\usepackage{colortbl}
\graphicspath{{./figs/}} % Make Latex search fig subdir for figures

\usepackage{amsmath} % Adds a large collection of math symbols
\usepackage{amssymb}
\usepackage{amsfonts}
\usepackage{upgreek} % Adds in support for greek letters in roman typeset

\newcommand*\patchAmsMathEnvironmentForLineno[1]{%
\expandafter\let\csname old#1\expandafter\endcsname\csname #1\endcsname
\expandafter\let\csname oldend#1\expandafter\endcsname\csname
end#1\endcsname
 \renewenvironment{#1}%
   {\linenomath\csname old#1\endcsname}%
   {\csname oldend#1\endcsname\endlinenomath}%
}
\newcommand*\patchBothAmsMathEnvironmentsForLineno[1]{%
  \patchAmsMathEnvironmentForLineno{#1}%
  \patchAmsMathEnvironmentForLineno{#1*}%
}
\AtBeginDocument{%
\patchBothAmsMathEnvironmentsForLineno{equation}%
\patchBothAmsMathEnvironmentsForLineno{align}%
\patchBothAmsMathEnvironmentsForLineno{flalign}%
\patchBothAmsMathEnvironmentsForLineno{alignat}%
\patchBothAmsMathEnvironmentsForLineno{gather}%
\patchBothAmsMathEnvironmentsForLineno{multline}%
\patchBothAmsMathEnvironmentsForLineno{eqnarray}%
}

\usepackage{hyperxmp}

\usepackage[pdftex,
            pdfauthor={\paperauthors},
            pdftitle={\paperasciititle},
            pdfkeywords={\paperkeywords},
            pdfcopyright={Copyright (C) \papercopyright},
            pdflicenseurl={\paperlicenceurl}]{hyperref}

\usepackage[all]{hypcap} % Internal hyperlinks to floats.

\usepackage{xspace} 
\usepackage{upgreek}

\def\lhcb {\mbox{LHCb}\xspace}

\def\velo   {VELO\xspace}

\def\MagUp {\mbox{\em Mag\kern -0.05em Up}\xspace}

\ifthenelse{\boolean{uprightparticles}}%
{

 \def\PDelta      {\ensuremath{\Delta}\xspace}                 
 \def\PXi      {\ensuremath{\Xi}\xspace}                 
 \def\PLambda      {\ensuremath{\Lambda}\xspace}                 
 \def\PSigma      {\ensuremath{\Sigma}\xspace}                 
 \def\POmega      {\ensuremath{\Omega}\xspace}                 
 \def\PUpsilon      {\ensuremath{\Upsilon}\xspace}                 
 
 \def\PB      {\ensuremath{\mathrm{B}}\xspace}                 
                  
 \def\PD      {\ensuremath{\mathrm{D}}\xspace}

 \def\PK      {\ensuremath{\mathrm{K}}\xspace}

 \def\Pb      {\ensuremath{\mathrm{b}}\xspace}                 
 \def\Pc      {\ensuremath{\mathrm{c}}\xspace}

 \def\Pi      {\ensuremath{\mathrm{i}}\xspace}

}
{

 \mathchardef\PDelta="7101
 \mathchardef\PXi="7104
 \mathchardef\PLambda="7103
 \mathchardef\PSigma="7106
 \mathchardef\POmega="710A
 \mathchardef\PUpsilon="7107
                  
 \def\PB      {\ensuremath{B}\xspace}                 
                  
 \def\PD      {\ensuremath{D}\xspace}

 \def\PK      {\ensuremath{K}\xspace}

 \def\Pb      {\ensuremath{b}\xspace}                 
 \def\Pc      {\ensuremath{c}\xspace}

 \def\Pi      {\ensuremath{i}\xspace}

}

\makeatletter
\ifcase \@ptsize \relax% 10pt
  \newcommand{\miniscule}{\@setfontsize\miniscule{4}{5}}% \tiny: 5/6
\or% 11pt
  \newcommand{\miniscule}{\@setfontsize\miniscule{5}{6}}% \tiny: 6/7
\or% 12pt
  \newcommand{\miniscule}{\@setfontsize\miniscule{5}{6}}% \tiny: 6/7
\fi
\makeatother

\DeclareRobustCommand{\optbar}[1]{\shortstack{{\miniscule (\rule[.5ex]{1.25em}{.18mm})}
  \\ [-.7ex] $#1$}}

\def\cquark    {{\ensuremath{\Pc}}\xspace}

\def\bquark    {{\ensuremath{\Pb}}\xspace}

  \def\Kbar    {{\kern 0.2em\overline{\kern -0.2em \PK}{}}\xspace}

\def\KorKbar    {\kern 0.18em\optbar{\kern -0.18em K}{}\xspace}

  \def\Dbar    {{\kern 0.2em\overline{\kern -0.2em \PD}{}}\xspace}

\def\DorDbar    {\kern 0.18em\optbar{\kern -0.18em D}{}\xspace}

\def\Bbar    {{\ensuremath{\kern 0.18em\overline{\kern -0.18em \PB}{}}}\xspace}

\def\BorBbar    {\kern 0.18em\optbar{\kern -0.18em B}{}\xspace}

  \def\Y#1S{\ensuremath{\PUpsilon{(#1S)}}\xspace}% no space before {...}!

\def\Lbar        {{\ensuremath{\kern 0.1em\overline{\kern -0.1em\PLambda}}}\xspace}
\def\LorLbar    {\kern 0.18em\optbar{\kern -0.18em \PLambda}{}\xspace}

\def\AT#1     {\ensuremath{A_{\mathrm{T}}^{#1}}\xspace}           % 2

\def\C#1      {\ensuremath{\mathcal{C}_{#1}}\xspace}                       % 9
\def\Cp#1     {\ensuremath{\mathcal{C}_{#1}^{'}}\xspace}                    % 7
\def\Ceff#1   {\ensuremath{\mathcal{C}_{#1}^{\mathrm{(eff)}}}\xspace}        % 9  
\def\Cpeff#1  {\ensuremath{\mathcal{C}_{#1}^{'\mathrm{(eff)}}}\xspace}       % 7
\def\Ope#1    {\ensuremath{\mathcal{O}_{#1}}\xspace}                       % 2
\def\Opep#1   {\ensuremath{\mathcal{O}_{#1}^{'}}\xspace}                    % 7

\newcommand{\tev}{\ifthenelse{\boolean{inbibliography}}{\ensuremath{~T\kern -0.05em eV}}{\ensuremath{\mathrm{\,Te\kern -0.1em V}}}\xspace}
\newcommand{\gev}{\ensuremath{\mathrm{\,Ge\kern -0.1em V}}\xspace}
\newcommand{\mev}{\ensuremath{\mathrm{\,Me\kern -0.1em V}}\xspace}
\newcommand{\kev}{\ensuremath{\mathrm{\,ke\kern -0.1em V}}\xspace}
\newcommand{\ev}{\ensuremath{\mathrm{\,e\kern -0.1em V}}\xspace}
\newcommand{\gevc}{\ensuremath{{\mathrm{\,Ge\kern -0.1em V\!/}c}}\xspace}
\newcommand{\mevc}{\ensuremath{{\mathrm{\,Me\kern -0.1em V\!/}c}}\xspace}
\newcommand{\gevcc}{\ensuremath{{\mathrm{\,Ge\kern -0.1em V\!/}c^2}}\xspace}
\newcommand{\gevgevcccc}{\ensuremath{{\mathrm{\,Ge\kern -0.1em V^2\!/}c^4}}\xspace}
\newcommand{\mevcc}{\ensuremath{{\mathrm{\,Me\kern -0.1em V\!/}c^2}}\xspace}

\def\mum  {\ensuremath{{\,\upmu\mathrm{m}}}\xspace}

\newcommand{\chisqip}{\ensuremath{\chi^2_{\text{IP}}}\xspace}

\def\gsim{{~\raise.15em\hbox{$>$}\kern-.85em
          \lower.35em\hbox{$\sim$}~}\xspace}
\def\lsim{{~\raise.15em\hbox{$<$}\kern-.85em
          \lower.35em\hbox{$\sim$}~}\xspace}

\def\ptot       {\mbox{$p$}\xspace}
\def\pt         {\mbox{$p_{\mathrm{ T}}$}\xspace}

\def\evtgen     {\mbox{\textsc{EvtGen}}\xspace}

\def\geant      {\mbox{\textsc{Geant4}}\xspace}

\def\photos     {\mbox{\textsc{Photos}}\xspace}

\def\pythia     {\mbox{\textsc{Pythia}}\xspace}

\def\tell1  {TELL1\xspace}
\def\ukl1   {UKL1\xspace}

\usepackage{cite} % Allows for ranges in citations
\usepackage{mciteplus}

\usepackage{longtable} % only for template; not usually to be used in PAPERs
\RequirePackage[normalem]{ulem} %DIF PREAMBLE
\RequirePackage{color}\definecolor{RED}{rgb}{1,0,0}\definecolor{BLUE}{rgb}{0,0,1} %DIF PREAMBLE
\providecommand{\DIFaddend}{} %DIF PREAMBLE
\begin{document}

%%%%%%%%%%%%%%%%%%%%%%%%%
%%%%% Title     %%%%%%%%%
%%%%%%%%%%%%%%%%%%%%%%%%%
\renewcommand{\thefootnote}{\fnsymbol{footnote}}
\setcounter{footnote}{1}

% %%%%%%% CHOOSE TITLE PAGE--------
%\onecolumn
%\input{title-LHCb-INT}
%\input{title-LHCb-ANA}
%\input{title-LHCb-CONF}
% $Id: title-LHCb-PAPER.tex 107952 2017-05-17 11:30:35Z uegede $
% ===============================================================================
% Purpose: LHCb-PAPER journal paper title page template
% Author: 
% Created on: 2010-09-25
% ===============================================================================

%%%%%%%%%%%%%%%%%%%%%%%%%
%%%%%  TITLE PAGE  %%%%%%
%%%%%%%%%%%%%%%%%%%%%%%%%
\begin{titlepage}
\pagenumbering{roman}

% Header ---------------------------------------------------
\vspace*{-1.5cm}
\centerline{\large EUROPEAN ORGANIZATION FOR NUCLEAR RESEARCH (CERN)}
\vspace*{1.5cm}
\noindent
\begin{tabular*}{\linewidth}{lc@{\extracolsep{\fill}}r@{\extracolsep{0pt}}}
\ifthenelse{\boolean{pdflatex}}% Logo format choice
{\vspace*{-2.7cm}\mbox{\!\!\!\includegraphics[width=.14\textwidth]{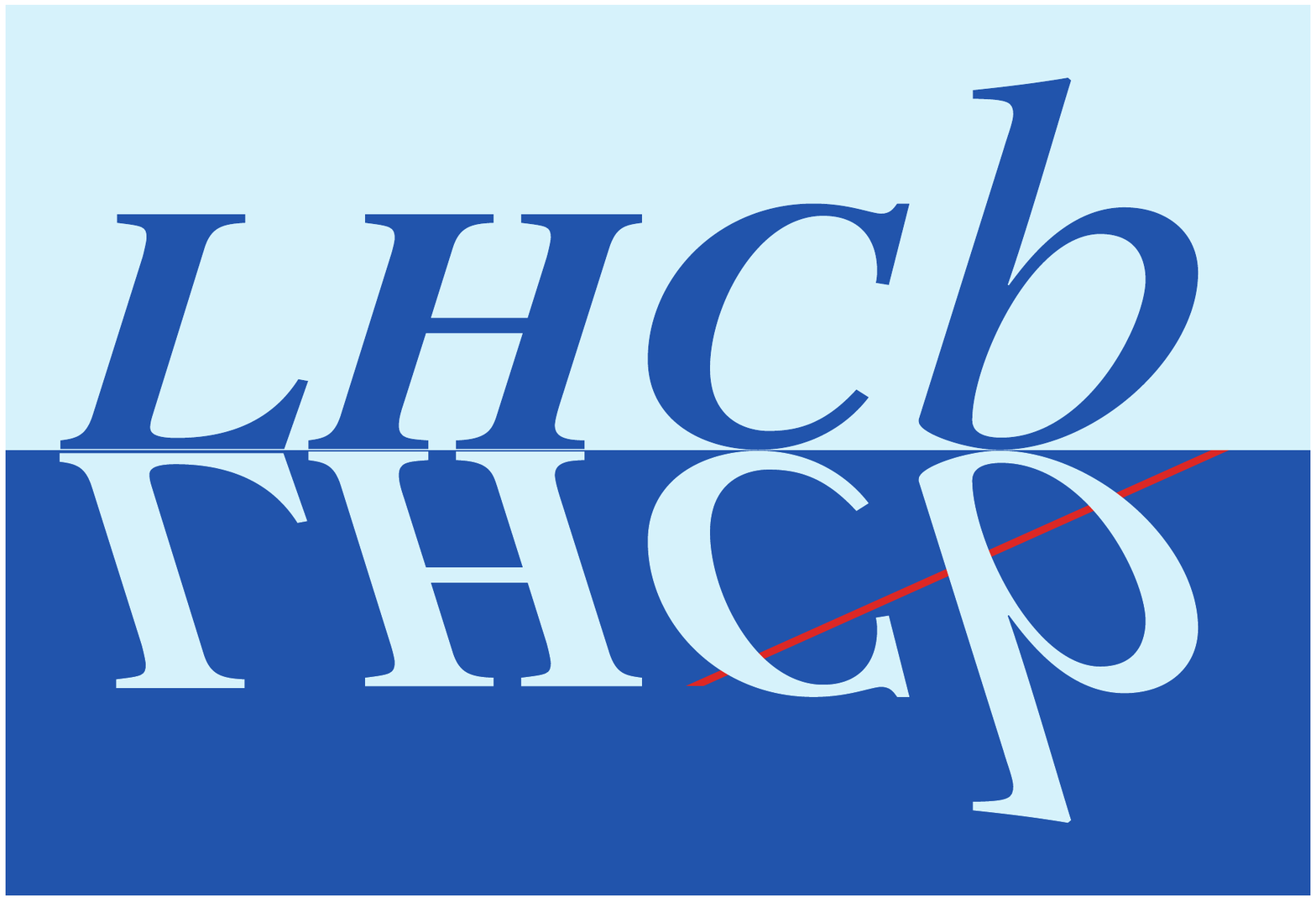}} & &}%
{\vspace*{-1.2cm}\mbox{\!\!\!\includegraphics[width=.12\textwidth]{lhcb-logo.eps}} & &}%
\\
 & & CERN-EP-2017-203 \\  % ID 
 & & LHCb-PAPER-2017-025 \\  % ID 
 & & September 5, 2017 \\ % Date - Can also hardwire e.g.: 23 March 2010
 & & \\
% not in paper \hline
\end{tabular*}

\vspace*{4.0cm}

% Title --------------------------------------------------
{\normalfont\bfseries\boldmath\huge
\begin{center}
% DO NOT EDIT HERE. Instead edit macro in main.tex to keep metadata correct
  \papertitle 
\end{center}
}

\vspace*{2.0cm}

% Authors -------------------------------------------------
\begin{center}
%In the footnote, replace 'paper' by 'Letter' in case of submission to PRL or PLB 
% Edit macro in main.tex to keep metadata correct
\paperauthors\footnote{Authors are listed at the end of this paper.}
\end{center}

\vspace{\fill}

% Abstract -----------------------------------------------
\begin{abstract}
  \noindent
Bose-Einstein correlations of same-sign charged pions, produced in proton-proton collisions at a 7~TeV centre-of-mass energy, are studied using a data sample collected by the LHCb experiment. The signature for Bose-Einstein correlations is observed in the form of an enhancement of pairs of like-sign charged pions with small four-momentum difference squared. The charged-particle multiplicity dependence of the Bose-Einstein correlation parameters describing the correlation strength and the size of the emitting source is investigated, determining both the correlation radius and the chaoticity parameter. The measured correlation radius is found to increase as a function of increasing charged-particle multiplicity, while the chaoticity parameter is seen to decrease.
\end{abstract}

\vspace*{1.0cm}

\begin{center}
  Published in JHEP 12 (2017) 025
\end{center}

\vspace{\fill}

{\footnotesize 
% Edit macro in main.tex to keep metadata correct
\centerline{\copyright~\papercopyright, licence \href{\paperlicenceurl}{\paperlicence}.}}
\vspace*{2mm}

\end{titlepage}

%%%%%%%%%%%%%%%%%%%%%%%%%%%%%%%%
%%%%%  EOD OF TITLE PAGE  %%%%%%
%%%%%%%%%%%%%%%%%%%%%%%%%%%%%%%%

%  empty page follows the title page ----
\newpage
\setcounter{page}{2}
\mbox{~}
%\newpage
%
%% Author List ----------------------------
%%  You need to get a new author list!
%\input{LHCb_authorlist.tex}
%
%The author list for journal publications is provided by the Membership Committee shortly after 'approval to go to paper' has been given.
%%It will be made available on the page
%%\verb!http://www.physik.uzh.ch/~strauman/forMemCo/LHCb-PAPER-XXXX-XXX/! .
%It will be sent to you by email shortly after a paper number has beens assigned.
%The author list should be included already at first circulation, 
%to allow new members of the collaboration to verify whether they have been included correctly.
%Occasionally a misspelled name is corrected or associated institutions become full members.
%In that case, a new author list will be sent to you.
%In case line numbering doesn't work well after including the authorlist, try moving the \verb!\bigskip! after the last author to a separate line.
%
%
%The authorship for Conference Reports should be ``The LHCb
%  collaboration'', with a footnote giving the name(s) of the contact
%  author(s), but without the full list of collaboration names.

\cleardoublepage

%\twocolumn
% %%%%%%%%%%%%% ---------

\renewcommand{\thefootnote}{\arabic{footnote}}
\setcounter{footnote}{0}

%%%%%%%%%%%%%%%%%%%%%%%%%%%%%%%%
%%%%%  Table of Content   %%%%%%
%%%%%%%%%%%%%%%%%%%%%%%%%%%%%%%%
%%%% Uncomment next 2 lines if desired
%\tableofcontents
%\cleardoublepage

%%%%%%%%%%%%%%%%%%%%%%%%%
%%%%% Main text %%%%%%%%%
%%%%%%%%%%%%%%%%%%%%%%%%%

\pagestyle{plain} % restore page numbers for the main text
\setcounter{page}{1}
\pagenumbering{arabic}

%% Uncomment during review phase. 
%% Comment before a final submission.
%DIF < %\linenumbers
%%\DIFaddbegin \linenumbers
\DIFaddend 

% You can include short sections directly in the main tex file.
% However, for larger papers it is desirable to split the text into
% several semiautonomous files, which can be revised independently.
% This is especially useful when developing a document in
% collaboration with several people, since then different parts can be
% edited independently.  This type of file organization is shown here.
% 

\section{Introduction}
\label{sec:Introduction}

Multiparticle production within the process of hadronisation can be investigated by measuring Bose-Einstein correlations (BEC) between indistinguishable bosons~\cite{BEC1,BEC3}. The technique to study the BEC effect in particle physics is the analogue of the Hanbury-Brown-Twiss (HBT) intensity interferometry~\cite{HBT1,HBT2,HBT3}. The production of identical bosons that are close in phase space is enhanced by the presence of BEC. The measurements of the quantum interference effect between indistinguishable particles emitted by a finite-size source are useful to understand the space-time properties of the hadron emission volume. 

Since the first observation of BEC in identically charged pions produced in $p \bar{p}$ collisions~\cite{Goldhaber}, the effect has been studied for multiboson systems produced in leptonic, hadronic and nuclear collisions~\cite{ALEPH1,ALEPH2,ALEPH3,ALEPH4,DELPHI1,DELPHI2,DELPHI3,L3_1,L3_2,L3_3,OPAL1,OPAL2,OPAL3,OPAL5,OPAL6,HERA_BEC,SPS_BEC,Tevatron_BEC,RHIC_BEC,ALICEpp1,ALICEpp2,CMSpp1,CMSpp2,ATLASpp,CMSpA,ALICEPbPb}. At the LHC, the BEC effect has been studied by the ALICE, ATLAS and CMS collaborations in proton-proton~\cite{ALICEpp1,ALICEpp2,CMSpp1,CMSpp2,ATLASpp}, proton-lead~\cite{CMSpA} and lead-lead~\cite{CMSpA,ALICEPbPb} collisions.

Dependences of the BEC effect upon various observables have been studied, including charged-particle multiplicity, average transverse momentum of the particle pair and boson mass. The latter has been reported by the LEP experiments~\cite{ALEPH1,ALEPH2,ALEPH3,ALEPH4,DELPHI1,DELPHI2,DELPHI3,L3_1,L3_2,L3_3,OPAL1,OPAL2,OPAL3,OPAL5,OPAL6}, and can be interpreted within some theoretical models~\cite{Alexander,BialasZalewski,BZKP1,BZKP2}.

In this paper, the first study of the BEC effect in $pp$ collisions in the forward region is presented. The BEC parameters characterising the correlation radius and the chaoticity of the correlation source are measured.

\section{BEC measurement}
\label{sec:BEC_Measurement}

Quantum interference effects are probed by studying the Lorentz invariant \mbox{quantity $Q$~\cite{BEC3,Baym}} of two indistinguishable particles of rest mass $m$ and four-momenta $q_{1}$ and $q_{2}$
\begin{linenomath}
\begin{equation}
\label{eqt:Q}
Q = \sqrt{-(q_1 - q_2)^2} = \sqrt{M^2 -4m^2},
\end{equation}
\end{linenomath}
which gives a measure of the phase-space separation of the two-particle system of invariant mass $M$.

\subsection{Two-particle correlation function}
\label{sec:Correlation_Function}

The BEC effect is expected to manifest itself as an enhancement in the two-particle correlation function in the low-$Q$ region below $\sim$0.5\gevcc, expressed as~\cite{Lorstad}
\begin{linenomath}
\begin{equation}
\label{eqt:Correlation_Function}
C_{2}(Q) = \frac{\rho_{2}(Q)}{\rho_{2}^{0}(Q)},
\end{equation}
\end{linenomath}
\noindent where $\rho_{2}(Q)$ is the two-particle density function for like-sign pairs of indistinguishable particles, as defined in Ref.~\cite{Lorstad}, and $\rho_{2}^{0}(Q)$ is the corresponding density function without the BEC effect, which is constructed as described in Sec.~\ref{sec:RefSamples}. The densities $\rho_{2}(Q)$ and $\rho_{2}^{0}(Q)$ are normalised to unity, such that they can be interpreted as probability density functions. The correlation function $C_2(Q)$ is commonly parameterised as a Fourier transform of the source density distribution, $C_2(Q) = N(1 + \lambda e^{-|R Q|^{\alpha_{\mathrm{L}}}})$~\cite{Levy}, where the parameter $R$, the correlation radius, can be interpreted as the radius of the spherically symmetric source of the emission volume, $N$ accounts for the overall normalisation and $\lambda$ is the chaoticity parameter, which accounts for the partial incoherence of the source~\cite{Deutschmann}. The chaoticity parameter can vary from zero, in the case of a completely coherent source, to unity for an entirely chaotic source. The Levy index of stability~\cite{Levy}, $\alpha_{\mathrm{L}}$, accounts for the assumed density distribution. The radial distribution of the static source corresponding to the case of $\alpha_{\mathrm{L}} = 1$ is used in the present analysis
\begin{linenomath}
\begin{equation}
C_2(Q) = N(1 + \lambda e^{-R Q}) \times (1 + \delta \cdot Q),
\label{eqt:CorrFunctLevy}
\end{equation}
\end{linenomath}
\noindent where the $\delta$ parameter accounts for long-range correlations, {\it e.g.} related to the transverse momentum conservation. This extended parameterisation follows better the $Q$ distribution in data, including in the low-$Q$ region below $\sim$0.5\gevcc~\cite{Kittel}.

The correlation function is, to first order, independent of the single-particle acceptance and efficiency. By construction of the correlation function, the effects due to the detector occupancy, acceptance and material budget are accounted for by dividing the $Q$ distribution for like-sign pion pairs by a reference distribution.

\subsection{Reference sample}
\label{sec:RefSamples}

The reference sample used to construct the $\rho_{2}^{0}(Q)$ density function, present in the denominator of Eq.~(\ref{eqt:Correlation_Function}), should reflect the distribution without the BEC effect while maintaining all other correlations. A number of reference samples can be constructed but none fully satisfies the above conditions. The reference sample may be constructed using experimental data, or with simulated events incorporating the detector interactions.

A data-driven ``event-mixed'' reference sample~\cite{Kopylov} is used in the present analysis. This approach is based on the choice of two identical bosons, each originating from different events, which naturally do not contain the BEC effect. However, this method of constructing boson pairs may not contain other correlations present in the same-sign boson data sample, such as correlations due to Coulomb interactions or long-range effects.

Alternative methods have been considered for constructing the reference sample. For example, the reference sample could consist of opposite-sign charged bosons originating from the same $pp$ interaction. As in the event-mixed reference sample, the main advantage of the opposite-sign approach is that the reference distribution is derived directly from data. However, the opposite-sign charge pairs may also originate from resonances which result in local enhancements in the $Q$ spectrum. Furthermore, correlations arising from the attraction of opposite charges are present in such a sample. Another method is to employ the simulated $Q$ distribution without the BEC effect. In this case, the crucial requirement is a good level of agreement between data and simulated samples in the distributions of crucial variables, {\it e.g.} the particle momenta. The absence of the Coulomb and spin effects in generators based on the Lund Model~\cite{Lund} may impinge on the correctness of this method.

\subsection{Double ratio}
\label{sec:DR}

To account for imperfections in the reference distribution derived from the data a ``double ratio'' $r_{\rm d}$ is commonly used in BEC studies
\begin{linenomath}
\begin{equation}
\label{eqt:DR}
r_{\rm d}(Q) \equiv \frac{C_{2}(Q)^{\mathrm{data}}}{C_{2}(Q)^{\mathrm{simulation}}},
\end{equation}
\end{linenomath}
\noindent where $C_{2}(Q)^{\mathrm{data}}$ denotes the correlation function in the data constructed using the event-mixed reference sample, while $C_{2}(Q)^{\mathrm{simulation}}$ indicates the correlation function in the simulation without the BEC effect, using an event-mixed sample built with simulated events in the same way as for data. The correlation function in the simulation without the BEC effect includes the simulated long-range correlations that are also present in data. Therefore, if the long-range correlations are correctly modelled, a constant $r_{\rm d}(Q)$ distribution is expected in the high-$Q$ region up to $\sim$2.0\gevcc. In the present analysis the BEC effect is measured by fitting the $r_{\rm d}(Q)$ distribution with the event-mixed reference sample, using the parameterisation given in Eq.~(\ref{eqt:CorrFunctLevy}).

\subsection{Coulomb correction}
\label{sec:Coulomb}

Final-state interactions involving both electromagnetic (Coulomb) and strong forces are present in the low-$Q$ region below $\sim$0.5\gevcc, and may potentially affect the distributions of the analysed observables. In the low-$Q$ region, the Coulomb repulsion between two identically charged hadrons alters the correlation function $C_{2}(Q)$ by decreasing the BEC effect. This effect is corrected for with the Gamov penetration factor~\cite{Coulomb1,Coulomb2}, $G_{2}(Q)$, by applying a weight per particle pair $1 / G_{2}(Q)$, where $G_{2}(Q) = \frac{2 \pi \zeta}{e^{2 \pi \zeta} - 1}$, $\zeta = \pm \frac{\alpha m}{Q}$, and $m$ and $\alpha$ denote the particle rest mass and the fine-structure constant, respectively. The sign of $\zeta$ is positive for same-charge and negative for opposite-charge pairs of hadrons.

The Coulomb interactions are not present in the simulated samples used in the analysis. This effect therefore has to be corrected for in the data.

\section{Detector and dataset}
\label{sec:Detector}

The \lhcb detector~\cite{Alves:2008zz} is a single-arm forward spectrometer designed for the study of particles containing \bquark or \cquark quarks. The detector includes a high-precision tracking system consisting of a silicon-strip \mbox{vertex detector (\velo)~\cite{LHCb-DP-2014-001}} surrounding the $pp$ interaction region and covering the pseudorapidity range $2<\eta <5$, a large-area silicon-strip detector located upstream of a dipole magnet with a bending power of about $4{\rm\,Tm}$, and three stations of silicon-strip detectors and straw drift tubes~\cite{LHCb-DP-2013-003} placed downstream of the magnet. The tracking system provides a measurement of momentum, \ptot,  with a relative uncertainty that varies from 0.5\% at low momentum to 1\% at $200$\gevc. The minimum distance of a track to a primary vertex (PV), the impact parameter (IP), is measured with a resolution of $(15+29/\pt)\mum$, where \pt is the component of \ptot transverse to the beam, in\gevc. Different types of charged hadrons are distinguished using information from two ring-imaging Cherenkov detectors~\cite{LHCb-DP-2012-003}. Photon, electron and hadron candidates are identified by a calorimeter system consisting of scintillating-pad and preshower detectors, an electromagnetic calorimeter and a hadronic calorimeter. Muons are identified by a system composed of alternating layers of iron and multiwire proportional chambers~\cite{LHCb-DP-2012-002}. The trigger~\cite{LHCb-DP-2012-004} consists of a hardware stage, based on information from the calorimeter and muon systems, followed by a software stage, which applies a full event reconstruction.

In the present analysis, a dataset of no-bias and minimum-bias triggered events collected in 2011 at a centre-of-mass energy of $\sqrt{s}$ = 7 TeV is used. The no-bias trigger selects events randomly, while the minimum-bias trigger requires at least one reconstructed \velo track. The data were collected with an average number of visible interactions per bunch crossing\footnote{A visible interaction corresponds to the PV reconstructed with at least five VELO tracks.} (pile-up) of 1.4~\cite{LHCb-PAPER-2014-047}. In order to eliminate biases related to the trigger requirements, a sample of ``independent $pp$ interactions'' is constructed as described in Sec.~\ref{sec:Selection}.

In the simulation, $pp$ collisions are generated using \pythia~8~\cite{Sjostrand:2007gs} with a specific \lhcb configuration~\cite{LHCb-PROC-2010-056} and without including the BEC effect.  Decays of hadronic particles are described by \evtgen~\cite{Lange:2001uf}, in which final-state radiation is generated using \photos~\cite{Golonka:2005pn}. The interaction of the generated particles with the detector and its response are implemented using the \geant toolkit~\cite{Allison:2006ve, *Agostinelli:2002hh}, as described in Ref.~\cite{LHCb-PROC-2011-006}. To study systematic effects, an additional sample is simulated using \pythia~6.4~\cite{Sjostrand:2006za} with the Perugia0 \cite{Perugia} tune.

\section{Selection and model fitting}
\label{sec:Selection}

The analysis uses a sample of events that may contain multiple $pp$ collisions. In the absence of trigger requirements each $pp$ interaction in the event can be analysed separately. Therefore, if the event is selected by the no-bias trigger, all PVs are accepted. In the case of events with multiple $pp$ collisions selected by the minimum-bias trigger, the related biases are suppressed by randomly removing one of the PVs containing the track(s) on which the trigger is fired.

The correlation function is constructed using pairs of same-sign pions. The particle identification (PID) is based on the output of a neural network employing subdetector information that quantifies the probability for a particle to be of a certain kind~\cite{LHCb-PROC-2011-008}. Such probabilities are calibrated to account for differences between data and the simulation that is used to train the neural network. The corrected values are derived from the data distributions using dedicated PID calibration samples~\cite{LHCb-DP-2012-003}. A high purity of the pion sample has to be ensured, but without suppressing low-momentum pions which mostly contribute to the signal region at low $Q$. The optimal limit on the pion identification probability is applied at the point where the signal enhancement in the low-$Q$ region below $\sim$0.5\gevcc for data begins to saturate. The pion purity with this selection remains high ($\sim$98\%). Additional vetoes on the kaon and proton identification probabilities are also imposed.

The following single particle requirements are applied. The selection requires that all pion candidates must have reconstructed track segments in the VELO, with $2 < \eta < 5$, and tracking stations downstream of the magnet. Each track must have a good-quality track fit, \pt~$> 0.1$\gevc, and no associated signal in the muon stations. Both pion candidates must be assigned to the same PV. Particles are assigned to the PV for which the $\chi^{2}$ value of the impact parameter, \chisqip, is the smallest, where \chisqip is defined as the difference in the vertex-fit $\chi^{2}$ of a given PV reconstructed with and without the track under consideration. A loose requirement on the track IP, IP~$<$~0.4~mm, is applied to retain most of the particles originating from a given PV. In order to reduce the contamination from fake and clone tracks,\footnote{Fake tracks are wrongly reconstructed tracks which combine the hits deposited by multiple particles in the tracking detectors. Clone tracks are two or more tracks reconstructed by mistake from the hits deposited in detectors by a single particle.} in the case where the tracks have all the same hits deposited in the VELO subdetector, only the track with the best $\chi^{2}$ is retained. In addition, fake tracks are removed using the requirements on the track $\chi^{2}$ and the output of a dedicated neural network~\cite{LHCb-PROC-2011-008}.

In the region $Q < 0.05$\gevcc, the separation in momentum between two particles is degraded and is not well simulated. The discrepancy between data and simulated track pairs tends to increase as $Q$ approaches zero. Investigations using simulation indicate that there is a significant fraction of pion pairs containing fake and clone tracks in the region $Q < 0.05$\gevcc for all activity classes. The double ratio is approximately constant and close to unity in the high-$Q$ region up to $Q \sim 2.0$\gevcc (see Fig.~\ref{fig:Fits_DR_pions}), which indicates that the long-range correlations are modelled accurately in this region. Consequently, the fits to the $r_{\rm d}$ distributions are restricted to the range $0.05 < Q < 2.0$\gevcc.

The BEC effect is expected to be largest in the low-$Q$ region below $\sim$0.5\gevcc, where it may be affected by same-sign clone tracks. Such clone pion pairs should manifest themselves as an enhancement in the distribution of the differences of the tangents of the track momenta of the two particles, where the tangents are measured in the $xz$ and $yz$ planes before the magnet, with the $z$ axis defined along the beam direction. The tangents are used to estimate the number of clone tracks remaining after the final selection, and the clone tracks can be suppressed with a requirement on the difference between the tangents of the two particles in a pair. Pion pairs are removed from the analysis if both $|\Delta t_{x}|$ and $|\Delta t_{y}|$ are less than 0.3~mrad, where $\Delta t_{x}$ and $\Delta t_{y}$ are the differences of the tangents of the track momenta of the two particles in the $xz$ and $yz$ planes. After applying these requirements, the effect of the clone particles is found to be negligible in the region $Q > 0.05$\gevcc.

\begin{figure}[tb]
\begin{center}
\includegraphics[width=0.7\linewidth]{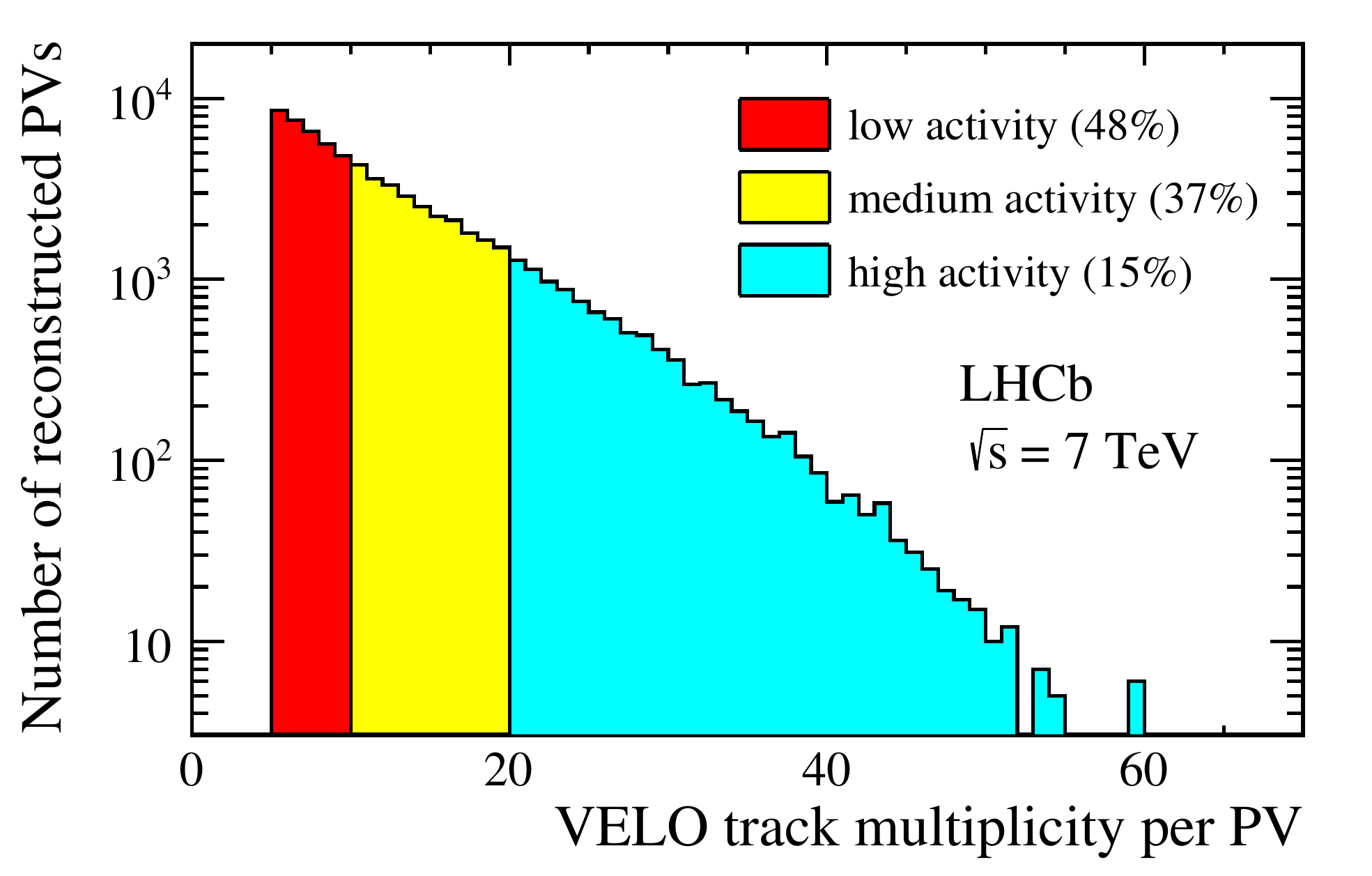}
\end{center}
\caption{Multiplicity of reconstructed \velo tracks assigned to a PV for the 2011 no-bias sample. Different colours indicate three activity classes defined as fractions of the full distribution. The minimum value of the track multiplicity to accept reconstructed PV is five.}
\label{fig:NchBins}
\end{figure}

The BEC parameters are studied as a function of the charged-particle multiplicity. However, the measured charged-particle multiplicities cannot be directly used to compare results among different experiments, mainly because the detector acceptances may not overlap and the reconstruction efficiencies may differ. This is why activity classes are introduced, reflecting the total multiplicity in the full solid angle. Three activity classes are defined in the range $2< \eta < 5$ according to the multiplicity of reconstructed VELO tracks assigned to a PV, which is a good probe of the total multiplicity. These activity classes are illustrated in Fig.~\ref{fig:NchBins}. The {\it low activity class} corresponds to a fraction of 48\% of PVs with lowest multiplicities (from 5 to 10 tracks). The {\it medium activity class} contains the 37\% of PVs with higher multiplicities (from 11 to 20 tracks). Finally, the {\it high activity class} contains 15\% of the highest multiplicity PVs ($\geq$~21 tracks). Using this classification, the comparison among different experiments is largely independent from specific features of the detectors. 

Although the activity classes have advantages in comparing results among various experiments characterised by different rapidity ranges, an unfolding procedure is performed to relate the reconstructed charged-particle multiplicities to those predicted by \pythia~8 with a specific \lhcb~configuration~\cite{LHCb-PROC-2010-056}. The multiplicity distributions are corrected using a Bayesian unfolding technique~\cite{Unfolding}. An unfolding matrix reflecting the probability of reconstructing a certain number of charged particles from a single PV in the range $2 < \eta < 5$ with generated charged-particle multiplicity $N_{ch}$ is populated using simulation and applied to the data. It is found that the corrected multiplicities agree well with the unfolded multiplicities previously determined by LHCb in Ref.~\cite{LHCb-PAPER-2013-070}. The activity classes correspond to the following generated charged-particle multiplicitiy intervals: $N_{ch} \in [8,18]$ (low activity), $N_{ch} \in [19,35]$ (medium activity) and $N_{ch} \in [36,96]$ (high activity).

The distributions of the double ratio of correlation functions in data and simulation for like-sign pion pairs, determined using the event-mixed reference sample, are fitted in the range $0.05 < Q < 2.0$\gevcc for the three different activity classes using the parameterisation of Eq.~(\ref{eqt:CorrFunctLevy}). The results of the binned maximum likelihood fit to the double ratio are summarised in Sec.~\ref{sec:Results}.

\section{Systematic uncertainties}
\label{sec:Systematics}

The properties of the correlation function and the construction of the double ratio make the fitted BEC parameters insensitive to the choice of the selection requirements to a large extent. However, due to imperfections in the reference sample and possible differences between data and simulation related to the generation model, as well as subtle reconstruction effects (like the reconstruction of close tracks sharing the same \velo hits or the track reconstruction in the high-occupancy detector regions), some second-order distortions in the double ratio may appear. The systematic uncertainties on the fit parameters, $R$ and $\lambda$, of the exponential model are determined by performing the analysis with modifications designed to estimate the systematic effects on individual contributions to the $r_{\rm d}(Q)$ distribution.

The leading source of systematic uncertainty is due to differences in the event generators used to determine the correlation function for the simulation. To study this effect, a sample of minimum-bias events produced using the \pythia~6.4 generator with Perugia0 tuning is used to construct the double ratio. The corresponding contribution to the systematic uncertainty is taken as the difference between the central values of the results obtained using the \pythia~8 and \pythia~6.4 datasets.

Another important source of systematic uncertainty is related to the PV multiplicity in the event. The constructed double ratio may be distorted in events containing multiple PVs, due to imperfections in the construction of the reference sample. To estimate the associated systematic uncertainty, the sample is divided into three subsamples containing events with one, two, and three or more PVs. For each subsample, the fit is performed and the maximum difference for each measured parameter is taken as a systematic uncertainty. The systematic uncertainty resulting from the PV reconstruction efficiency is also considered. To account for the effect of pile-up in the data and inefficiencies in the PV reconstruction, a systematic uncertainty is estimated as the difference between the nominal fit results and the results obtained from a fit to the data in which the PV reconstruction has been repeated after removing randomly a subset of the tracks from the event.

After applying the track quality requirements, the fraction of remaining fake tracks is determined from simulation to be at the level of 1\%. To determine the systematic uncertainty due to the presence of such fake tracks, the double ratio is refitted with looser track quality requirements. A similar uncertainty is obtained from a second method in which sets of randomly selected uncorrelated tracks are added. The observed change in BEC parameters is negligible with respect to the statistical uncertainty.

The fraction of like-sign pion pairs containing a clone track after the selection is determined to be below 1\%. The systematic uncertainty due to the presence of clone tracks is estimated by fitting the double ratio $r_{\rm d}(Q)$, after applying a tight requirement on the Kullback-Leibler distance~\cite{Kullback} such that the clone contribution is fully removed in simulation. The effect is found to be negligible for all activity classes.

The systematic uncertainty due to the calibration of the particle identification in the simulation is estimated by comparing several variants of the calibration procedure with the acceptance evaluated in different binning schemes for the particle momentum, pseudorapidity and track multiplicity. The largest difference after refitting the double ratios is taken as a systematic uncertainty. 

As the requirement on the pion identification probability alters the contamination of pions due to misidentification, it can influence the values of the $R$ and $\lambda$ parameters. The contribution of this effect to the systematic uncertainty is estimated by refitting $r_{\rm d}(Q)$ with the requirement on the pion identification probability changed to increase the fraction of misidentified pions by 50\%.

The systematic uncertainty derived from the fit range in the low-$Q$ (high-$Q$) region is determined by changing the lower (upper) limit of the $Q$ value by $\pm$0.01\gevcc ($\pm$0.2\gevcc). The fits to the double ratio with two different lower (upper) limits of $Q$ are performed for the three activity classes and the largest difference is taken as a systematic uncertainty.

The systematic uncertainty due to Coulomb corrections is estimated by varying the corrections by $\pm$20\%. The variation in the fit parameters is found to be less than 0.1\%, and is therefore neglected. It is also found that imposing different requirements on the particle IPs has no significant influence on the measured correlation radius or chaoticity parameter. The fractions of kaon-kaon and proton-proton like-sign pairs misidentified as a pion pair in the pion sample in the BEC signal region of $Q < 1.0$\gevcc are found to be negligible. Pairings of different particle types have a negligible effect.

Other effects like the fit binning, the resolution of the $Q$ variable, different magnet polarities, beam-gas interactions and residual acceptance effects related to possible differences between data and simulation in the low-$Q$ region below $\sim$0.2\gevcc, are also studied and found to be negligible.

The contributions to the systematic uncertainty are listed in Table~\ref{tab:Systematics}. Correlations of the systematic uncertainties between different activity classes are negligible.

\begin{table}[]
\caption{Fractional systematic uncertainties on the $R$ and $\lambda$ parameters for the three activity classes, as described in the text. The total uncertainty is the sum in quadrature of the individual contributions.}
\begin{center}\begin{tabular}{lccccccc}
\hline
Source & \multicolumn{2}{c}{Low activity} & \multicolumn{2}{c}{Medium activity} & \multicolumn{2}{c}{High activity} \\
\hline
& $\Delta R$ [\%] & $\Delta \lambda$ [\%] & $\Delta R$ [\%] & $\Delta \lambda$ [\%] & $\Delta R$ [\%] & $\Delta \lambda$ [\%] \\
\hline
 Generator tunings       & 6.6 & 4.3 & ~8.9 & 3.5 & 6.5 & 1.5 \\
 PV multiplicity         & 5.9 & 5.8 & ~6.1 & 4.5 & 3.9 & 4.3 \\
 PV reconstruction       & 1.8 & 0.1 & ~1.4 & 1.2 & 0.1 & $<$0.1~~~\\
 Fake tracks             & 0.4 & 1.1 & ~1.7 & 3.9 & 1.1 & 0.8 \\
 PID calibration         & 1.3 & 0.3 & ~0.8 & 0.6 & 2.7 & 0.9 \\
 Requirement on pion PID & 2.9 & 1.8 & ~1.6 & 0.1 & 1.3 & 0.1 \\
 Fit range at low-$Q$    & 1.2 & 1.0 & ~1.2 & 1.5 & 1.8 & 2.7 \\
 Fit range at high-$Q$   & 1.8 & 0.1 & ~2.1 & 0.8 & 2.4 & 1.4 \\
\hline
 Total                   & 9.8 & 7.6 & 11.4 & 7.3 & 8.8 & 5.6 \\
\hline
\end{tabular}\end{center}
\label{tab:Systematics}
\end{table}

\section{Results}
\label{sec:Results}

\begin{figure}[H]
\begin{center}
\includegraphics[width=0.60\linewidth]{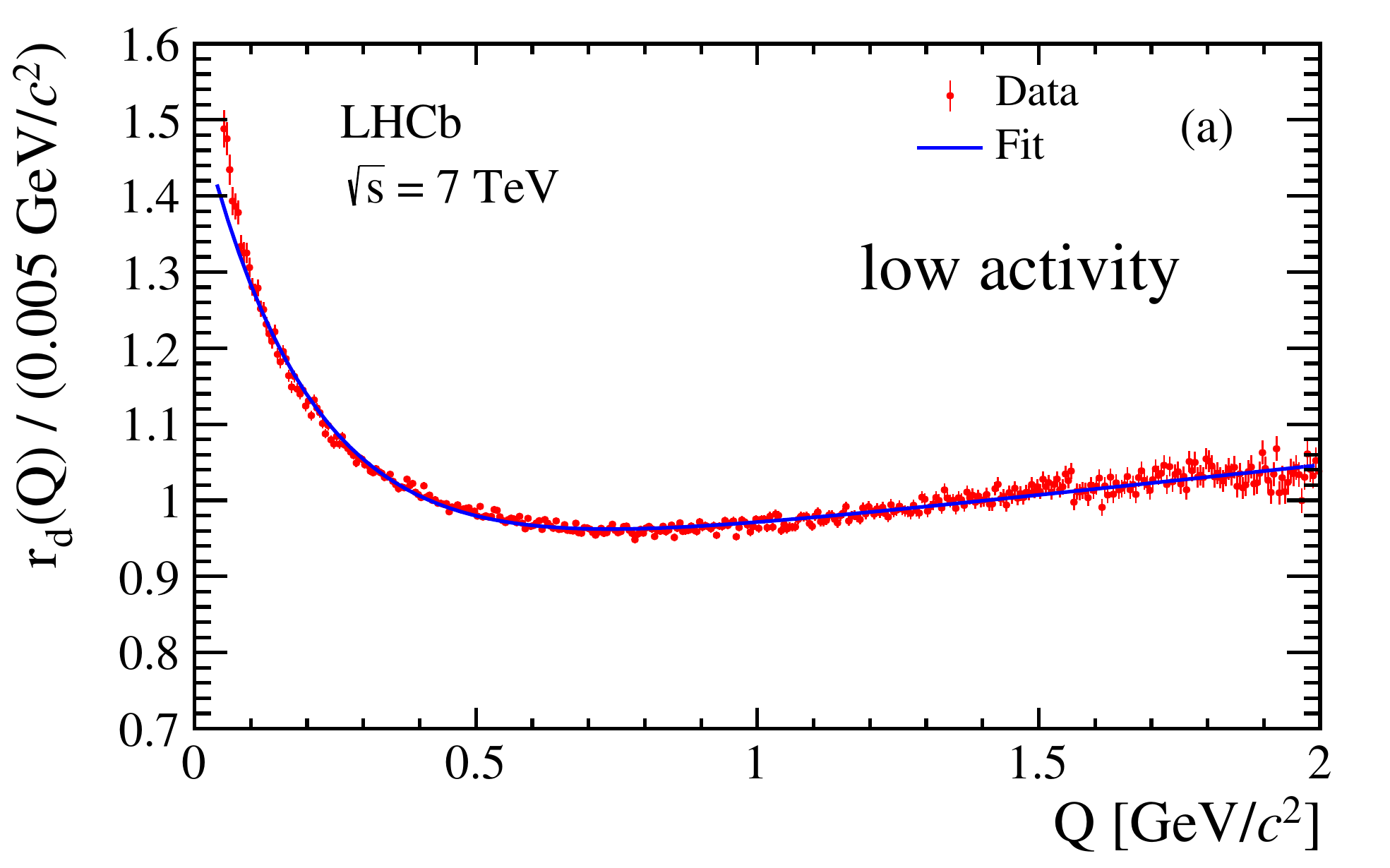}
\includegraphics[width=0.60\linewidth]{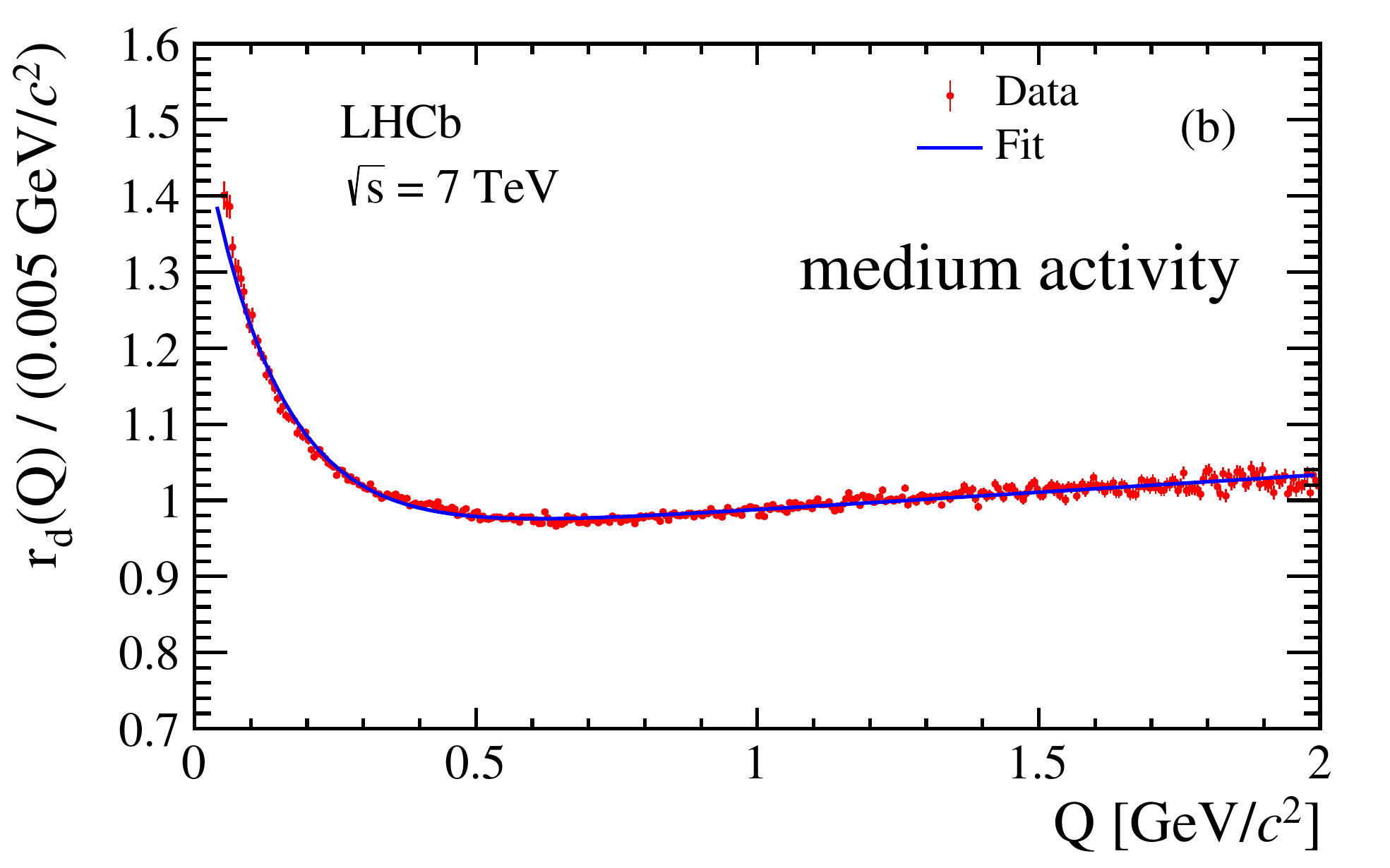}
\includegraphics[width=0.60\linewidth]{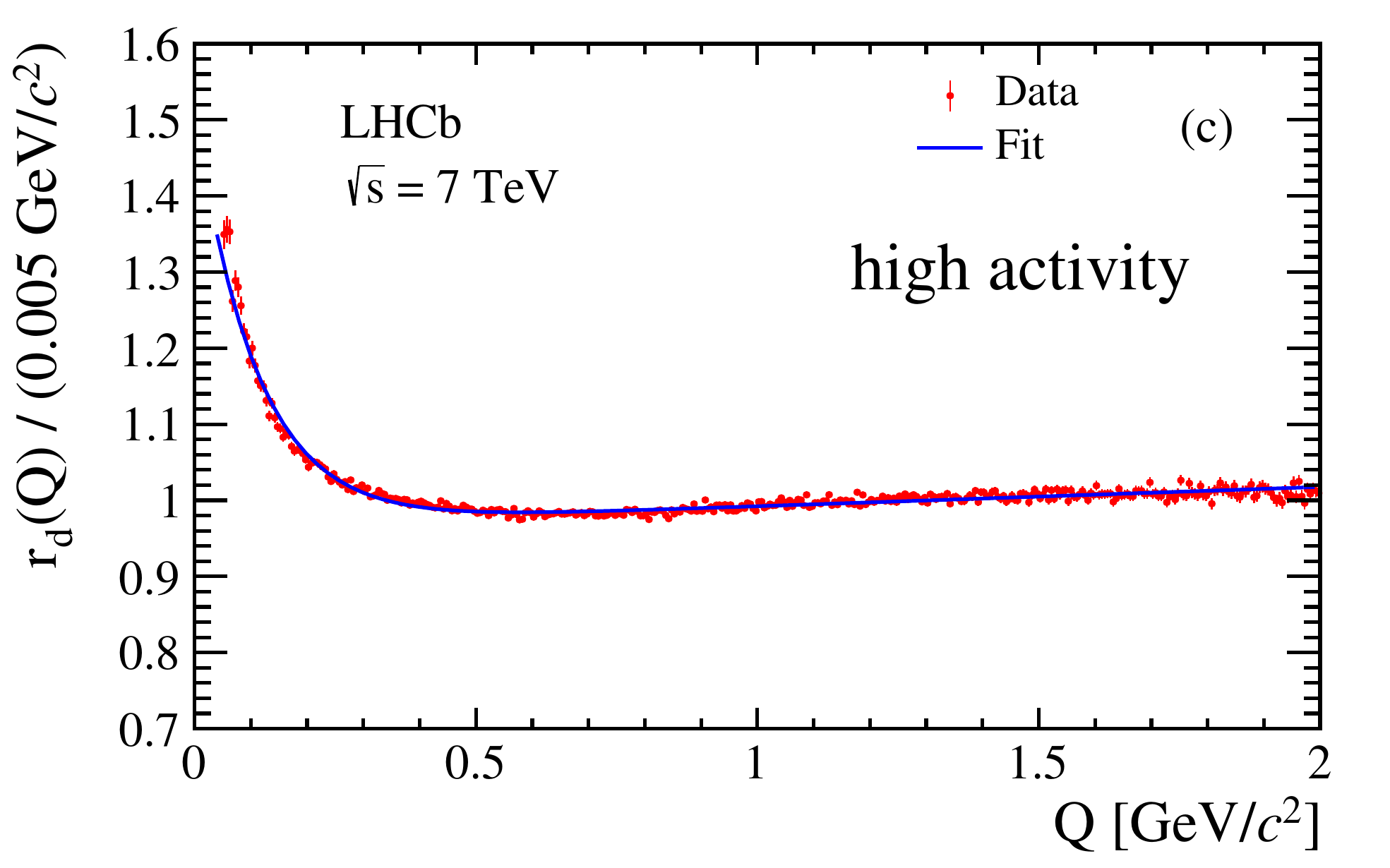}
\end{center}
\caption{Results of the fit to the double ratio for like-sign pion pairs with event-mixed reference samples and the Coulomb effect subtracted for the three activity classes: (a) low, (b) medium and (c) high activity. The blue solid line denotes the fit result using the parameterisation of Eq.~(\ref{eqt:CorrFunctLevy}). Only statistical uncertainties are shown.}
\label{fig:Fits_DR_pions}
\end{figure}

The results of fits to the double ratios for the correlation radius, chaoticity parameter and $\delta$ parameter for the three different activity classes are summarised in Table~\ref{tab:Fits_DR_pions}, including statistical and systematic uncertainties, and are presented in Fig.~\ref{fig:Fits_DR_pions}.

\begin{figure}[H]
\begin{center}
\includegraphics[width=0.7\linewidth]{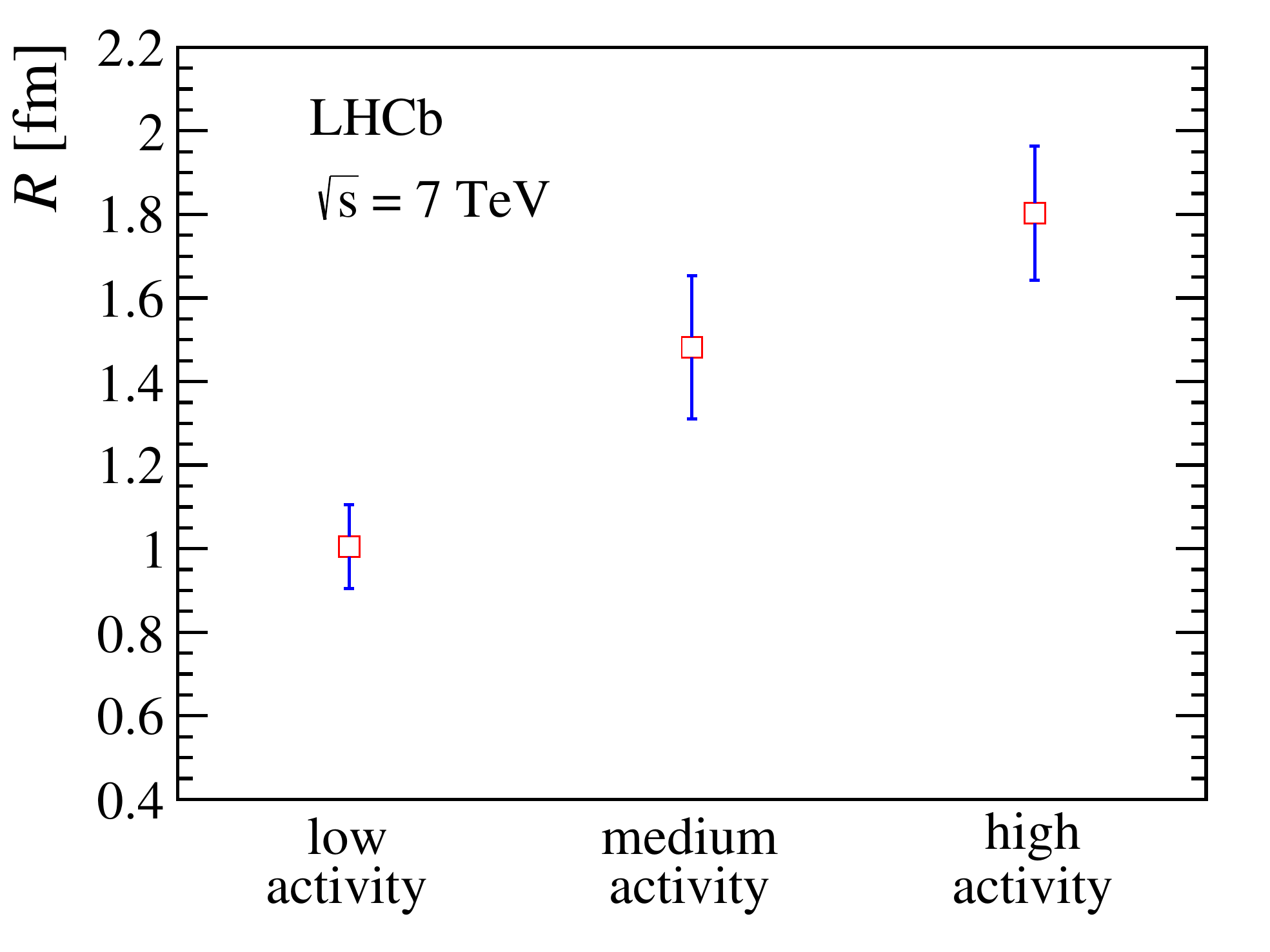}
\end{center}
\caption{Correlation radius $R$ as a function of activity. Error bars indicate the sum in quadrature of the statistical and systematic uncertainties. The points are placed at the centres of the activity bins.}
\label{fig:depR}
\end{figure}

\begin{figure}[H]
\begin{center}
\includegraphics[width=0.7\linewidth]{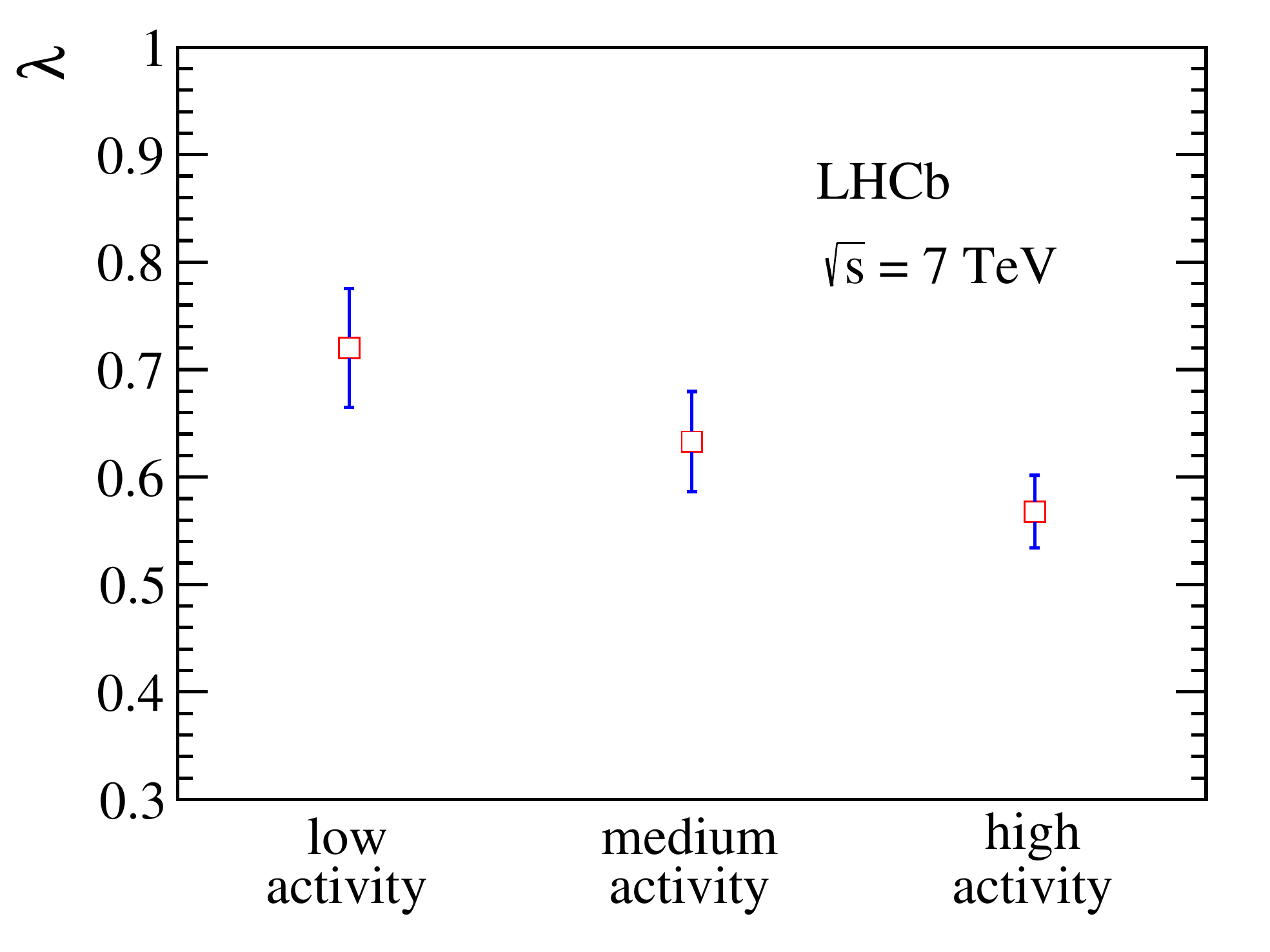}
\end{center}
\caption{Chaoticity parameter $\lambda$ as a function of activity. Error bars indicate the sum in quadrature of the statistical and systematic uncertainties. The points are placed at the centres of the activity bins.}
\label{fig:depL}
\end{figure}

The dependences of the correlation radius and the chaoticity parameter on the activity class are shown in Figs.~\ref{fig:depR} and~\ref{fig:depL}, respectively. As the activity class increases, the $R$ parameter also increases, while the $\lambda$ parameter decreases. This confirms previous observations at LEP~\cite{OPAL3} and in the other LHC experiments~\cite{ALICEpp1,CMSpp1,CMSpp2,ATLASpp}. There are no theoretical predictions for the BEC effect in $pp$ interactions, however the observed trends are qualitatively predicted within some theoretical models~\cite{Kittel,Suzuki,Buschbeck,Sarkisyan}.

Due to the different pseudorapidity coverage of LHCb with respect to other LHC experiments, the comparison of the measured BEC parameters for a given multiplicity out of a $pp$ interaction is not straightforward. In the case of unfolded multiplicities in different pseudorapidity ranges quoted by experiments, the correspondence can be found using relations obtained from simulated events. The results for $pp$ collisions at 7~TeV published by the ATLAS experiment~\cite{ATLASpp} are quoted for unfolded multiplicities in the pseudorapidity range $|\eta| < 2.5$ and \pt~$> 0.1$\gevc. \pythia~8 is used to determine the relation for the multiplicity bins defined in the LHCb ($2 < \eta < 5$) and ATLAS ($|\eta| < 2.5$ and \pt~$> 0.1$\gevc) acceptances. The data indicate that the LHCb results for both $R$ and $\lambda$ are slightly below the ATLAS ones at 7~TeV. In order to perform a more detailed comparison it would be necessary to measure the BEC parameters using a full three-dimensional analysis~\cite{BEC3D}. 

It should be noted that the fit quality using the parameterisation, Eq.~(\ref{eqt:CorrFunctLevy}), is poor (see Fig.~\ref{fig:Fits_DR_pions}). The $\chi^2$ values are equal to 591, 623 and 621 for 386 degrees of freedom for low, medium and high activity classes, respectively. The difference between the fitted function and the data points, visible in the whole $Q$ range, is particularly large in the low-$Q$ BEC signal region below 0.2\gevcc. This indicates that the approximate parameterisation of Eq.~(\ref{eqt:CorrFunctLevy}) does not reproduce the measured distribution properly. Such an effect is observed also by other experiments~\cite{CMSpp2,ATLASpp}. This may introduce an additional systematic uncertainty in the theoretical interpretation of the fit results.

\begin{table}[]
\caption{Results of fits to the double ratio $r_{\rm d}(Q)$ for the three different activity classes and corresponding $N_{ch}$ bins, using the parameterisation of Eq.~(\ref{eqt:CorrFunctLevy}). Statistical and systematic uncertainties are given separately.}
\begin{center}\begin{tabular}{lcccc}
\hline
Activity & $N_{ch}$ & $R$ [fm] & $\lambda$ & $\delta$ [GeV$^{-1}$]\\
\hline
Low    & [8,18]   & 1.01 $\pm$ 0.01 $\pm$ 0.10   & 0.72 $\pm$ 0.01 $\pm$ 0.05   & 0.089 $\pm$ 0.002 $\pm$ 0.044\\
Medium & [19,35]  & 1.48 $\pm$ 0.02 $\pm$ 0.17   & 0.63 $\pm$ 0.01 $\pm$ 0.05   & 0.049 $\pm$ 0.001 $\pm$ 0.009\\
High   & [36,96]  & 1.80 $\pm$ 0.03 $\pm$ 0.16   & 0.57 $\pm$ 0.01 $\pm$ 0.03   & 0.026 $\pm$ 0.001 $\pm$ 0.010\\
\hline
\end{tabular}\end{center}
\label{tab:Fits_DR_pions}
\end{table}

\section{Summary and conclusions}

Using a data sample collected by the LHCb experiment in proton-proton collisions at a centre-of-mass energy of 7~TeV, the Bose-Einstein correlations between two indistinguishable pions are studied in the forward acceptance region of $2 < \eta < 5$ for single pions with transverse momentum \pt~$> 0.1$\gevc. An enhancement of pairs of same-sign charged pions with small relative momentum related to the BEC effect is observed. An event-mixed reference sample is used to determine the signal and the double ratio distributions are fitted using an exponential parameterisation. The results confirm that the effective size of the emission region increases as a function of increasing charged-particle multiplicity, while the chaoticity parameter decreases, as previously observed at LEP and at the other LHC experiments. The $R$ and $\lambda$ parameters measured in the forward region in three different charged-particle multiplicity bins are slightly lower with respect to those measured by ATLAS for corresponding $pp$ interaction multiplicities.

\newpage

\section*{Acknowledgements}

\noindent We express our gratitude to our colleagues in the CERN
accelerator departments for the excellent performance of the LHC. We
thank the technical and administrative staff at the LHCb
institutes. We acknowledge support from CERN and from the national
agencies: CAPES, CNPq, FAPERJ and FINEP (Brazil); MOST and NSFC (China);
CNRS/IN2P3 (France); BMBF, DFG and MPG (Germany); INFN (Italy); 
NWO (The Netherlands); MNiSW and NCN (Poland); MEN/IFA (Romania); 
MinES and FASO (Russia); MinECo (Spain); SNSF and SER (Switzerland); 
NASU (Ukraine); STFC (United Kingdom); NSF (USA).
We acknowledge the computing resources that are provided by CERN, IN2P3 (France), KIT and DESY (Germany), INFN (Italy), SURF (The Netherlands), PIC (Spain), GridPP (United Kingdom), RRCKI and Yandex LLC (Russia), CSCS (Switzerland), IFIN-HH (Romania), CBPF (Brazil), PL-GRID (Poland) and OSC (USA). We are indebted to the communities behind the multiple open 
source software packages on which we depend.
Individual groups or members have received support from AvH Foundation (Germany),
EPLANET, Marie Sk\l{}odowska-Curie Actions and ERC (European Union), 
Conseil G\'{e}n\'{e}ral de Haute-Savoie, Labex ENIGMASS and OCEVU, 
R\'{e}gion Auvergne (France), RFBR and Yandex LLC (Russia), GVA, XuntaGal and GENCAT (Spain), Herchel Smith Fund, The Royal Society, Royal Commission for the Exhibition of 1851 and the Leverhulme Trust (United Kingdom).

\newpage

\addcontentsline{toc}{section}{References}
\setboolean{inbibliography}{true}
\bibliographystyle{LHCb}
\bibliography{BEC,main,LHCb-PAPER,LHCb-CONF,LHCb-DP,LHCb-TDR}

\newpage

% Author List ----------------------------                                                                                                                                                                                                                                                                                                
%  You need to get a new author list!                                                                                                                                                                                                                                                                                                    

\centerline{\large\bf LHCb collaboration}
\begin{flushleft}
\small
R.~Aaij$^{40}$,
B.~Adeva$^{39}$,
M.~Adinolfi$^{48}$,
Z.~Ajaltouni$^{5}$,
S.~Akar$^{59}$,
J.~Albrecht$^{10}$,
F.~Alessio$^{40}$,
M.~Alexander$^{53}$,
A.~Alfonso~Albero$^{38}$,
S.~Ali$^{43}$
% ...

{\footnotesize \it
% ...
$ ^{5}$Clermont Universit{\'e}, Universit{\'e} Blaise Pascal, CNRS/IN2P3, LPC, Clermont-Ferrand, France\\
% ...
$ ^{10}$Fakult{\"a}t Physik, Technische Universit{\"a}t Dortmund, Dortmund, Germany\\
% ...
$ ^{38}$ICCUB, Universitat de Barcelona, Barcelona, Spain\\
$ ^{39}$Universidad de Santiago de Compostela, Santiago de Compostela, Spain\\
$ ^{40}$European Organization for Nuclear Research (CERN), Geneva, Switzerland\\
% ...
$ ^{43}$Nikhef National Institute for Subatomic Physics, Amsterdam, The Netherlands\\
% ...
$ ^{48}$H.H. Wills Physics Laboratory, University of Bristol, Bristol, United Kingdom\\
% ...
$ ^{53}$School of Physics and Astronomy, University of Glasgow, Glasgow, United Kingdom\\
% ...
$ ^{59}$University of Cincinnati, Cincinnati, OH, United States\\
% ...
}
\end{flushleft}

%\newpage
%\input{LHCb_authorlist.tex}

%The author list for journal publications is generated from the Membership Database shortly after 'approval to go to paper' has been given.
%It will be sent to you by email shortly after a paper number has been assigned.
%The author list should be included already at first circulation,
%to allow new members of the collaboration to verify that they have been included correctly.
%Occasionally a misspelled name is corrected, or associated institutions become full members.
%Therefore an updated author list will be sent to you after the final EB review of the paper.
%In case line numbering doesn't work well after including the authorlist, try moving the \verb!\bigskip! after the last author to a separate line.

%The authorship for Conference Reports should be ``The LHCb collaboration'', with a footnote giving the name(s) of the contact
%  author(s), but without the full list of collaboration names.

\end{document}